\documentstyle[12pt]{article}

\begin{document}

\topmargin 0pt
\oddsidemargin 5mm

\setcounter{page}{1}
\vspace{2cm}
\begin{center}

{\bf Finite size effects in Derrida's model multicriticities and
limits of generalization for the Zamolodchikov' s C-theorem}\\
\vspace{5mm}
{\large D. Saakian}\\
\vspace{5mm}
{\em Yerevan Physics Institute}\\
{Alikhanian Brothers St.2, Yerevan 375036, Armenia\\
and
Laboratory of Computing techniques and automation, JINR,
41980 Dubna, Russia\\
Saakian @ vx1.YERPHI.AM
\\
$PACS.05.50 + q$ - Lattice theory and statics, Ising problems\\
$PACS.75.10Nr$ - Spin - Glass and other random models}
\end{center}

\vspace{5mm}
\centerline{\bf{Abstract}}
Finite size effects in the multicriticity point and boundaries between phases
 are calculated.  There are anomalous large finite size
effects on the boundary of ferromagnetic phase with paramagnetic or spin-glass.
Multicriticity point is not giving global minimum for the finite size
corrections of free energy.


\vspace{5mm}

According to $C$-theorem of Zamolodchikov[1], one can introduce some function
$c(g)$ for $2d$ conformal theories, which decreases along the renorm-group
paths. Kutasov [2] has suggested to consider as a $c(g)$ partition for the
models on the $2d$ gravitation. For some model he found, that partition has
maximum at multicriticity points. \\
The question arose, is it general property for any phase transition?\\
Independent of $C$ theorem some years ago I began to investigate another
(similar!) problem [3]. If symmetry was broken spontanously during the phase
transition and system has chosen some vacuum from the set, it should spend
some free energy-"cost of decision making". To found it, we need in
asymptotics for the $\ln Z(N,T)$, where $Z$ is a partition for a system of
$N$ spins at temperature $T$. Let we can define in some manner
\begin{equation}
\lim_{N\to \infty}\ln Z(N,T)=F(N,T)+o(1)
\end{equation}
Then at phase transition point $T_c$ function F(N,T) probably has some jump,
describing waste of free energy for vacuum choosing:
\begin{equation}
\lim_{\epsilon \to 0}F(N,T_c-\epsilon)-F(N,T_c-\epsilon)\ne 0
\end{equation}
If we have some expansion for the $F(N,T)$,
\begin{equation}
F(N,T)=f_0(T)N+..  f_1 \ln N+f_2(T)+o(1)
\end{equation}
then jump should be due to term $f_2(T)$ as a rule, sometimes due to $f_1(T)$.
It was well known, that for the case of 2d Ising model on a periodic rigid
lattice jump equals $\ln 2$. For the $2d$ Potts models with $Z(Q)$ symmetry and
$3d$ Ising model Monte-Carlo simulations give [3], that jump for all this models
on the rigid periodic lattices equals $\ln Q$. For the Ising model on
the dynamical lattices jump equals $g\ln 2$ [4], where g is a genus number of
surface. Perhaps such proportionality between jump and $g$ absens for a
nonunitar models[5].\\
A question arose to find this jumps for a general case of phase transitions. It
is also interesting to calculate $F(N,T)$ at phase transition point with
accuracy inclusively $O(1)$. Such problem was well known in statistical
mechanics [6-7], where correction terms are defined by conformal symmetry. \\
We are going to calculate free energy of REM with finite size corrections, to
understand jumps of $F$. It is mean-field like simple system, which could be
solved with finite size corrections. REM [8] is a system, with the independent
 gaussian distrubution of $2^N$ energy levels. For a single ferromagnetic
 energy level we have a distribution [9]
\begin{equation}
P_1(E_1)=\frac{1}{\sqrt{\pi N}}\exp(-(E_1+J_0N)^2/N)
\end{equation}
and for other $2^N-1$ levels
\begin{equation}
P(E)=\frac{1}{\sqrt{\pi N}}\exp(-E^2/N)
\end{equation}
For the free energy we have an expression [9,10]
\begin{equation}
\label{A1}
<\ln Z>=\Gamma'(1)+\int_{-\infty}^{\infty}ud[f(u+u_f)f(u)^M]
\end{equation}
where $u_f=J_0NB, \lambda=B\sqrt{N},M=2^N-1$ and
\begin{equation}
\label{A2}
f(u)=\frac{1}{2\pi i}\int_{-i\infty}^{i\infty}\Gamma(x)
\exp[-ux+\lambda^2x^2/4]dx
\end{equation}
In this integral the loop over passes point $0$ from right.
Function $f(u)$ is monotonic, like step. With exponential accuracy it equals $1$
below $0$, then become $0$ above it (with the same accuracy). We need in three
asymptotic regimes
\begin{equation}
\label{a3}
\begin{array}{ll}
f(u)\approx\frac{1}{\sqrt{\pi}\lambda}\Gamma (2u/\lambda^2)\exp(-u^2/\lambda^2),
 \lambda\ll u \\
f(u)\approx\frac{1}{\sqrt{\pi}}\int_{u/\lambda}^{\infty}dx\exp(-x^2),
 \lambda\ll \mid u \mid \ll \lambda^2 \\
f(u)\approx 1-\frac{1}{\sqrt{\pi}\lambda}\Gamma (-2u/\lambda^2)\exp(-u^2/\lambda^2),
-\lambda^2/2<u\ll-\lambda\\
f(u)\approx 1-exp(u+\lambda^2/4),
-\lambda^2<u<-\lambda^2/2
\end{array}
\end{equation}
As the $[f(u+u_f)f(u)^M$ likes step function, its derivative is like $\delta$
function with a coordinate of wall at some $-u_0$. The vicinity of that point
gives main contribution to the integral in (6) (bulk value is $u_0$).
Ferromagnetic phase appears, when wall of function $f(u+u_f)$ is lefter, than
wall of $f(u)^M$.
For the  derivative we derive
\begin{equation}
\label{AA3}
f'(u)=-\exp[u+\lambda^2/4]f(u+\lambda^2/2)
\end{equation}
For our convinent presentation of $f(u)$ it is easy to calculate value of
$f(u),f^{'}(u)$ at 2 points:
\begin{equation}
\label{AA4}
\begin{array}{ll}
f(0)=1/2,&f^{'}(0)=-\frac{1}{\lambda \sqrt{\pi}}\\
f(-\lambda^2/2)=1-1/2\exp[-\lambda^2/4],
&f^{'}(-\lambda^2/2)=-1/2\exp[-\lambda^2/4]
\end{array}
\end{equation}
As was found in [8], in paramagnetic phase $O(1)$ corrections to free energy
 disappear, situation is the same in the ferromagnetic phase [9].
In the case of SG phase there are corrections $-\frac{B}{2B_c}\ln N$ [8].\\
Let us consider boundary SG-PM. In this case we can neglect by
ferromagnetic level. We have $U_0=\lambda^2/2$. The boundary described by a
line
\begin{equation}
\label{A4}
\begin{array}{ll}
0<J_0< \sqrt{\ln 2}\\
B=\sqrt{\ln 2}
\end{array}
\end{equation}
Near the $u=-u_0$ we consider an expansion by degrees of $u+u_0$ for $f(u)$
like
\begin{equation}
\label{A5}
f(u)=1-a\exp[-\lambda^2/4+b(u+u_0)+..]
\end{equation}
We found, that
\begin{equation}
\label{A6}
f(u)=1-\frac{1}{2}\exp[u+\lambda^2/4][1+O(\frac{1}{\lambda})]
\end{equation}
Let us take $M$ power of (13)
\begin{equation}
\label{A7}
\exp[-\exp(u+u_0-\ln 2)]=\exp(-\phi)
\end{equation}
Its solution
\begin{equation}
\label{A8}
u=-u_0+\ln 2+\ln \phi
\end{equation}
So (6) goes to
\begin{equation}
\label{A9}
<\ln Z>=\Gamma^{'}(1)-\int_{-\infty}^{\infty}ue^{-\phi}d\phi=\\
\Gamma^{'}(1)-u_0-[\ln 2+\Gamma^{'}(1)]=\\
2N\ln 2-\ln 2
\end{equation}
Let us consider boundary between FM and SG phases. It is a line
\begin{equation}
\label{A17}
\begin{array}{ll}
J_0=\sqrt{\ln 2}\\
\infty>B>\sqrt{\ln 2}
\end{array}
\end{equation}
We using the property, that if there is only 1-st level, $<\ln Z>=J_0NB$ :
\begin{equation}
\label{A18}
\Gamma'(1)+\int_{-\infty}^{\infty}ud[f(u+u_f)]=J_0NB
\end{equation}
Using this equality, after simple transformations we derive
\begin{equation}
\label{A19}
\begin{array}{ll}
<\ln Z>=\Gamma'(1)+\int_{-\infty}^{\infty}ud[f(u+u_f)f(u)^M]\\
=\Gamma'(1)+\int_{-\infty}^{\infty}udf(u+u_f)
-\int_{-\infty}^{\infty}ud[f(u+u_f)[1-f(u)^M]\\
=J_0NB+\int_{-\infty}^{\infty}f(u+u_f)[1-f(u)^M]du\\
\end{array}
\end{equation}
We have a product of 2 monotonic functions, decreasing (one-to left, another-
to right) far the point $u=-u_f$.
Let us introduce $F(u)$
\begin{equation}
\label{A20}
F^{'}(u)=f(u+u_f)
\end{equation}
At $ \lambda\ll \mid u \mid \ll \lambda^2$ we derive
\begin{equation}
\label{A21}
F(u-u_f)=\int_{0}^{u/\lambda}dx
[\frac{\lambda}{\sqrt{\pi}}\int_{x}^{\infty}\exp [-t^2]dt
-\frac{C}{\sqrt{\pi}}\exp(-u^2/\lambda^2)]
\end{equation}
After transformations $<\ln Z>$ goes to
\begin{equation}
\label{A22}
\begin{array}{ll}
<\ln Z>&=\Gamma'(1)+\int_{-\infty}^{\infty}ud[f(u+u_f)f(u)^M]\\
&=\Gamma'(1)+\int_{-\infty}^{\infty}udf(u+u_f)
-\int_{-\infty}^{\infty}ud[f(u+u_f)[1-f(u)^M]\\
&=J_0NB+\int_{-\infty}^{\infty}[f(u+u_f)[1-f(u)^M]du
\end{array}
\end{equation}
Let $\Psi(u)=1-f(u)^M$. Then
\begin{equation}
\label{A23}
\begin{array}{ll}
<\ln Z>&=J_0NB+\int_{-\infty}^{\infty}F^{'}(u)\Psi(u)du\\
&=J_0NB+F(\infty)\Psi(\infty)-F(-\infty)\Psi(-\infty)
-\int_{-\infty}^{\infty}F(u)\Psi^{'}(u)du\\
&=J_0NB+\int_{-\infty}^{\infty}[F(\infty)-F(u)]\Psi^{'}(u)du\\
&=J_0NB+F(\infty)-F(-u_f)
-F^{'}(-u_f)\int_{-\infty}^{\infty}(u+u_f)\Psi^{'}(u)du
\end{array}
\end{equation}
We truncated expansion in degrees $u+u_f$, because $\Psi^{'}(u)$ is
similar to  $\delta$ function near the $-u_f$. Then we derive
\begin{equation}
\label{A24}
<\ln Z>=J_0NB+
\frac{\lambda}{\sqrt{\pi}}\int_{0}^{\infty}dx\int_{x}^{\infty}\exp [-t^2]dt-C/2
-\frac{1}{2}\int_{-\infty}^{\infty}(u+u_f)\Psi^{'}(u)du
\end{equation}
  Let us take last integral using representation $d\Psi(u)=-e^{-\phi}d\phi$.
We have
\begin{equation}
\label{A25}
\begin{array}{ll}
e^{-\phi}=\frac{A}{\lambda}\exp[-U^2/\lambda^2+N\ln2]\\
\ln\phi=-U^2/\lambda^2+N\ln2+\ln A-\ln \lambda
\end{array}
\end{equation}
Let us consider expansion $u=u_f+u_1$. For the $u_1$ we derive an equation
\begin{equation}
\label{A26}
u_1=(\ln\phi-\ln A+\ln \lambda)B/B_c,
B_c=2\sqrt{\ln{2}}
\end{equation}
Eventually we have an expression
\begin{equation}
\label{A27}
<\ln Z>=J_0NB+
\frac{B\sqrt{N}}{\sqrt{\pi}}\int_{0}^{\infty}dx\int_{x}^{\infty}\exp [-t^2]dt
-C/2-[\ln N+ \Gamma'(1)+\ln\frac{B\sqrt{\pi}}{\Gamma(\frac{B_c}{B})}]B/(2B_C)
\end{equation}
We see the $\sqrt{N}$ order corrections, which are very strange.\\
For the boundary PM-FM we have an equation
\begin{equation}
\label{A28}
\begin{array}{ll}
J_0B=B^2/4+\ln 2\\
J_0>\sqrt{\ln 2}
\end{array}
\end{equation}
Everthing is similar to previous section, only
\begin{equation}
\label{A29}
\Psi(u)=\exp -[\exp(u+u_0)],u+u_0=\ln \phi
\end{equation}
For free energy we derive
\begin{equation}
\label{A30}
<\ln Z>=J_0NB+
\frac{B\sqrt{N}}{\sqrt{\pi}}\int_{0}^{\infty}dx\int_{x}^{\infty}\exp [-t^2]dt
-\Gamma^,(1)/2
\end{equation}
So $\sqrt{N}$ corrections stay.
At multicriticity point everything is the same, as in previous section, only
\begin{equation}
\label{A31}
\Psi(u)=\exp\{ -\frac{1}{2}[\exp(-\lambda^2/4+u+u_f)]\}
\end{equation}
So
\begin{equation}
\label{A32}
\begin{array}{ll}
u+u_f=\ln \phi+\ln 2\\
<\ln Z>=J_0NB+
\frac{B\sqrt{N}}{\sqrt{\pi}}\int_{0}^{\infty}dx\int_{x}^{\infty}\exp [-t^2]dt
-C/2-(\Gamma'(1)+\ln 2)/2
\end{array}
\end{equation}
We see, that on FM-PM line $<\ln Z>$ is more, than at tricritical point.
Perhaps it is connected with the lack of unitarity in spin glass models
(or unhomogenuity [10]).
This work was supported by  German ministry of Science and Technology
Grant 211 - 5231.

\end{document}